\documentclass{ws-ijgmmp}
\begin{document}

\markboth{L. G. Molinari, C.A. Mantica}
{w=1/3 to w=1 evolution in a RW space-time}

%
\catchline{}{}{}{}{}
%

\title{w=1/3 TO w=-1 EVOLUTION\\ IN A ROBERTSON-WALKER SPACE-TIME\\ WITH CONSTANT SCALAR CURVATURE}

\author{LUCA GUIDO MOLINARI}
\address{Dipartimento di Fisica ``A. Pontremoli'',
Universit\`a degli Studi di Milano\\ and INFN sez. di Milano,
Via Celoria 16, I-20133 Milano, Italy\\
\email{luca.molinari@unimi.it}
}
\author{CARLO ALBERTO MANTICA}
\address{I.I.S. Lagrange, Via L. Modignani 65, 
I-20161, Milano, Italy \\
and INFN sez. di Milano,
Via Celoria 16, I-20133 Milano, Italy\\
\email{carlo.mantica@mi.infn.it}
}

\maketitle

\begin{history}
\received{(Day Month Year)}
\revised{(Day Month Year)}
\end{history}

\begin{abstract}
The Ricci tensor of a Robertson-Walker space-time is here specified by requiring constancy 
of the scalar curvature and a vanishing spatial curvature. By entering this Ricci tensor in Einstein's
equations (without cosmological constant), the cosmological fluid shows a transition from a pure radiation to 
a Lambda equation of state. 
In other words, the RW geometry with constant scalar curvature and flat space, fixes the 
limit values w=1/3 and w=-1, without any hypothesis on the cosmological fluid. 
The value of the scalar curvature fixes the time-scale for the transition. 

For this reason, we investigate the `toy-universe' with Hubble parameter $h=0.673$ and temperature $T_{CMB}=2.72$K. 
The model predicts an age of the universe in the range 7.3-13.7 Gyr.
\end{abstract}

\section{Introduction}
The spectral properties of the cosmic microwave background~\cite{Planck} and the galaxy surveys~\cite{Guzzo} 
firmly support the cosmological principle. 
Therefore, the Robertson-Walker (RW) geometry of space-time is the stage 
for large-scale cosmology.
In the standard $\Lambda$CDM model, the Friedmann-RW equations are solved for a mixture of radiation, baryonic and cold dark matter, 
and dark energy described by the cosmological constant. Their equations of state 
consistently determine the time evolution of geometry and densities. Fundamental physics at the different
energy scales determines the transitions from one era to another. In the present cold era,
radiation, matter and dark-energy only interact via the geometry. \\
A remarkable yet inexplicable fact is the perduring flatness of space, $R^*=0$, since the first inflation, at
the very beginning of space-time \cite{Weinberg,Longair}.

In this work we fix the geometry of a RW space-time by
requiring a null spatial curvature, $R^*=0$, and a constant positive scalar curvature $R$. 
We build the Ricci tensor in a covariant way, by solving for its eigenvalues (sections 2 and 3). 
Only then (section 4), the geometry is entered in the Einstein equations, without cosmological constant, 
and the properties of the cosmological fluid are read out. The outcome is at first intriguing: 
the cosmological fluid shows a pressure/density ratio evolving from the value 1/3 to the value -1,
i.e. the values of standard cosmology.\\
For this only reason, in section 5 we explore this toy-universe with cosmological data that are most
model-independent: the Hubble parameter and the temperature of the microwave
background. The solution imposes a dark energy as a form of missing energy of gravitational nature,
different from a cosmological term. The toy-universe produces an age of the present universe in a range 7.3-13.7 
Gyr. depending on the fraction radiation/ gravitational dark energy at $t=0$. In any case the ratio is very small
to ensure that all energy densities are positive.

\section{A covariant description of RW space-times}
A RW space-time is characterised by a preferred frame, with warped metric 
\begin{align}
 ds^2 = -dt^2 + a(t)^2 g^*_{\mu\nu} (x) dx^\mu dx^\nu \label{metric}
 \end{align}
$a(t)>0$ is the scale function and $g^*(x)$ is the metric tensor of a maximally symmetric 
Riemannian sub-manifold (latin index is for space-time, greek is
for space). To study the geometry, we introduce a covariant description.

RW space-times belong to a hierarchy that includes generalised RW space-times (GRW) \cite{Mantica:2017C} and
Twisted space-times. In twisted space-times, $g^*$ is the metric tensor of a Riemannian sub-manifold, and 
$a^2$ may depend also on space coordinates. In GRW space-times $g^*$ is still general, a Riemannian metric, but
$a^2$ may only depend on time. These spaces can be defined covariantly.

In any dimension, a
twisted space-time is characterised by the existence of a velocity field $u_k$ ($u^ku_k=-1$) that
is shear-free, vorticity-free and acceleration-free (i.e. torse-forming): 
\begin{align}
\nabla_iu_j=\varphi (u_iu_j+g_{ij}) \label{torse}
\end{align}
where $\varphi $ is a scalar field \cite{Mantica:2017B}.
The space-time is GRW if $\nabla_j\varphi= -u_j\dot\varphi $ (a dot means a 
directional derivative $u^k\nabla_k$). For RW space-times
the further requirement is that the Weyl tensor
is zero \cite{Mantica:2017B}. These results of us were inspired by the works
by Bang-Yen Chen on warped manifolds \cite{Chen:2014,Chen_torqued,Chen:2017}.

For these space-times, we obtained the general form of the Ricci tensor \cite{Mantica:2016C,Mantica:2016A}.
In particular, the Ricci tensor of a four-dimensional RW space-time is 
\begin{align}
R_{ij}=\tfrac{1}{3}(R -4\xi)u_i u_j +\tfrac{1}{3}(R-\xi)g_{ij} \label{Ricci}
\end{align}
where $R$ is the scalar curvature and $\xi $ is the eigenvalue $R_{ij}u^j =\xi u_i$, with 
\begin{align}
\xi = 3 (\dot\varphi + \varphi^2).
\end{align} 
This expression results from the evaluation of $R_{ij}u^j = [\nabla_m,\nabla_i]u^m$ with \eqref{torse}
and the condition $\nabla_i\varphi = -u_i\dot\varphi $.\\
In the locally comoving frame ($u_0=-1$, $u_\mu=0$), the metric has the form 
\eqref{metric} and:
\begin{align}
\varphi (t) = \frac{\dot a}{a}=H, \quad \xi(t) =3\frac{\ddot a}{a}=3(H^2+\dot H)
\end{align}
where $H(t)$ is Hubble's function. 
The covariant divergence of the equation $R_{ij}u^j=\xi u_i$ and the identity $\nabla^i R_{ij} =\frac{1}{2}\nabla_j R$, 
give 
\begin{align}
\dot R -2\dot\xi = -2\varphi (R -4\xi) \label{dotRxi}
\end{align}
The equation has the solution
\begin{align}
R - 2\xi = \frac{R^*}{a^2} + 6\varphi^2  \label{Rsol}
\end{align}
where the integration constant $R^*$ is identified with the constant curvature of the space-like 
sub-manifold, in the co-moving reference frame.

\section{RW geometry for constant $\mathbf R$ and $\mathbf{R^*=0}$}
If $R$ is constant, eq.\eqref{dotRxi} for the eigenvalue can be integrated. It
depends on the cosmological time through the scale function:
\begin{align}
\xi(t) = \frac{R}{4} \left[ 1- \frac{A}{a(t)^4} \right]  \label{xia}
\end{align}
where $A$ is a constant. Eq.\eqref{Rsol} with $R^*=0$ (k$=0$ in cosmology) becomes: $R=6 (\dot a/a)^2 + 2\xi (t) $.
The two equations give 
$$12 (\dot a)^2 = R[a^2 + Aa^{-2} ]$$
If $a(t)=ct^\gamma + ... $ for small $t$, the equation shows that
$12\gamma^2 c^2 =RA/c^2$ and $2(\gamma -1)=-2\gamma $. Then 
$\gamma =1/2$ and, for $R>0$, it is $A>0$. Put $a(t) = A^{1/4}\sqrt {x(t)}$ and obtain $(3/R) \dot x^2 = x^2 + 1 $. Take the 
square root with the choice $\dot x>0$ (expansion). Integration with the initial
condition $a(0)=0$ gives:
\begin{align}
a(t) = A^{1/4} \sqrt{\sinh \theta} , \qquad \theta = t\sqrt{\tfrac{R}{3}}    \label{at}
\end{align} 
The logarithmic derivative in $t$ gives Hubble's parameter:
\begin{align}
H (t) \equiv \frac{\dot a}{a} = \tfrac{1}{2} \sqrt{\tfrac{R}{3}} \coth \theta \label{Hubble}
\end{align}
The deceleration parameter is
\begin{align}
 q (t)\equiv - \frac{a\ddot a}{\dot a^2} = -\frac{\xi}{3\varphi^2} = 1-2\tanh^2 \theta \label{dec}
\end{align}
It changes from 1 to -1, and changes sign at $\bar\theta = \log (\sqrt 2 +1)\approx 0.881$. 
With $H (t) $ and $q(t)$ we obtain the constant curvature scalar: 
\begin{align}
R= 6[1-q(t)] H^2(t)   \label{R}
\end{align}

\section{The cosmological fluid}
The Einstein equations $R_{ij} - \frac{1}{2}R g_{ij} =\kappa T_{ij}$ ($\kappa =8\pi G$) relate the Ricci tensor \eqref{Ricci}
to the energy-momentum density tensor of a perfect-fluid, 
\begin{align*}
\tfrac{1}{3}(R-4\xi) u_iu_j - \tfrac{1}{6}(R+2\xi) g_{ij} =\kappa (p+\rho )v_iv_j + \kappa p g_{ij}
\end{align*}
By transvecting with a space-like vector orthogonal to $u_j$ and $v_j$ one gets the pressure
\begin{align}
\kappa p = -\tfrac{1}{6}(R+2\xi) 
\end{align}
The equality $\tfrac{1}{3}(R-4\xi) u_iu_j =\kappa (p+\rho )v_iv_j $ is transvected with $u^j$. Since 
$u^jv_j$ cannot vanish (or $v_j$ would be space-like), it must be $v_j$ collinear with $u_j$. The 
energy density  of the cosmological fluid is
\begin{align}
\kappa \rho = \tfrac{1}{2}R - \xi
\end{align}
If we now impose the pre-determined geometry \eqref{xia}, the fluid's density and pressure are:
\begin{align}
& \rho (t) =  \tfrac{R}{4\kappa} \coth^2 \theta \\
&p (t) = \tfrac{R}{12\kappa } \coth^2 \theta -\tfrac{R}{3\kappa } 
\end{align}
The ratio of $p/\rho $ is the function plotted in Figure 1:
\begin{align}
\frac{p}{\rho} = \frac{1}{3} -\frac{4}{3} \tanh^2\theta
\end{align}
The pressure is zero at $\theta^*=\frac{1}{2}\log 3\approx 0.549$, when $H(t^*) = \sqrt{R/3}$, $q(t^*)=0.5$. \\
For $\theta \ll 1$ both $\rho $ and $p$ diverge, with ratio $p /\rho \to \frac{1}{3}$. \\
For $\theta \gg 1$: $\rho (t) \approx \tfrac{R}{4\kappa} $ and $p(t)\approx -\tfrac{R}{4\kappa }$ i.e. $p/\rho \to -1$. \\
In real cosmology, the limit $1/3$ would correspond to a radiation era, while 
the  limit $-1$ would
correspond to the $\Lambda $ or dark-energy era, with a negative pressure that imparts a 
positive acceleration to the expansion.
\begin{figure}
\begin{center}
\includegraphics[width=7cm,clip=]{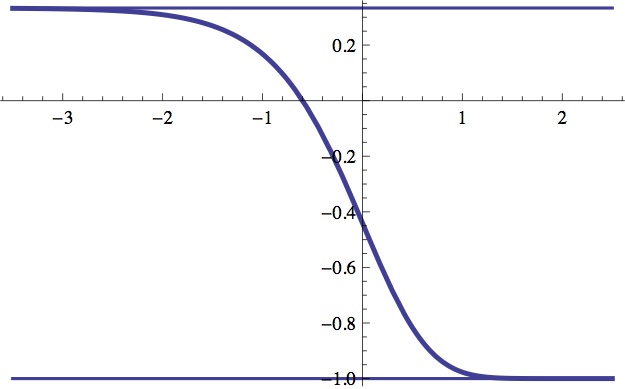}
\caption{The ratio $p/\rho$ as a function of  $\ln \theta$.  The asymptotic values are $p/\rho =\frac{1}{3}$ for $\theta \ll 1$
and $p/\rho =-1$ for $\theta \gg 1$ (Mathematica).}
\end{center}
\end{figure}

\section{A toy-model}
Being $R^*=0$, the total energy density $\rho (\theta )$ is critical all the time.
Let's assume that the fluid is composed
of radiation (R), with equation of state $p_R =\frac{1}{3}\rho_R$,
matter (M) with $p_M =0$, and a form of `dark energy' due to gravity 
with density $\rho_G$ and pressure $p_G$. Their sums are:
\begin{align}
&\rho_R(\theta ) + \rho_M (\theta ) + \rho_G (\theta )=\rho (\theta) \label{EV1}\\
&\tfrac{1}{3}\rho_R(\theta) + p_G (\theta) = p(\theta ) \label{EV2}
\end{align}
If $t_0$ is the present time (with parameter $\theta_0$) we introduce the present ratios 
$\Omega_x =\rho_x(\theta_0)/\rho (\theta_0)$. It is $\Omega_R +\Omega_M+\Omega_G=1$.

Planck's data \cite{Planck} come in two values:
`Planck only' and (Planck and baryonic acoustic oscillations). We read:
\begin{align*}
\Omega_M  = 0.3158 &\quad (0.3106)  \\
\Omega_\Lambda  = 0.6842 &\quad (0.6894)  \\
H_0 = 67.32 &\quad (67.70)\quad \text{(km/s Mpc)}
\end{align*}
Since $\Omega_M+\Omega_\Lambda =1$, today's radiation component is negligible and, today, 
$\Omega_G=\Omega_\Lambda $.\\
With the value $H_0$ we evaluate today's critical energy density (the total energy density in this model):
$$ \rho (\theta_0) =  \frac{3H_0^2}{\kappa} =  8.51319 \times 10^{-27} \text{kg/m}^3 $$
The cosmic microwave radiation (CMB) provides accurate model-independent information. The radiation 
energy density is the sum of a CMB term and a neutrino term. The first one is evaluated with the present 
CMB temperature $T_0=2.7255$~K: $\Omega_\gamma =5.45\times 10^{-5}$. 
The neutrino term is $\Omega_\nu = \tfrac{7}{8} N_\nu^{eff} (4/11)^{4/3} \Omega_\gamma = 0.6813\; \Omega_\gamma $ \cite{Weinberg}. Then 
\begin{align}
\Omega_R   = 9.16\times 10^{-5} \label{OMEGAR}
\end{align}
Radiation and matter evolve with scaling laws:
\begin{align}
\rho_R (\theta ) = \rho_R(\theta_0) \frac{a(\theta_0)^4}{a(\theta)^4}, \quad
\rho_M (\theta ) =\rho_M(\theta_0) \frac{a(\theta_0)^3}{a(\theta)^3}  \label{scalinglaws}
\end{align}
Both densities vanish at large times, while the total density $\rho (\theta )$ has a finite limit. 
Then $\rho_G (\theta) \to R/4\kappa $ and $p_G (\theta ) \to -R/4\kappa $ for $\theta\gg 1$. \\
For $\theta \to 0$ radiation dominates matter and diverges with power $1/\theta^2$, as the
total energy density. Since nothing is known about $\rho_G$, we may assume that also the dark energy scales as $1/\theta^2 $ at small times:
$$  \rho_G (\theta) \approx \frac{R}{4\kappa} \frac{1-\alpha}{\theta^2} $$ 
where $0<\alpha <1$ is a parameter. We shall see that a very small but nonzero $\alpha $ is necessary
to ensure $\rho_G\ge 0$ at all times. Now, for small $\theta$: 
\begin{gather*}
\rho_R (\theta_0) \frac{\sinh^2\theta_0}{\theta^2} + \frac{R}{4\kappa} \frac{1-\alpha }{\theta^2} = \frac{R}{4\kappa }\frac{1}{\theta^2} \quad\text{i.e.} \\
\rho_R (\theta_0)\sinh^2\theta_0 = \frac{R}{4\kappa }\alpha
\end{gather*}
As a consequence: 
\begin{align}
&\Omega_R \cosh^2\theta_0 =\alpha \label{CON1}\\
&\rho_R(\theta ) = \frac{R}{4\kappa} \frac{\alpha }{\sinh^2\theta}  \label{CON2}
\end{align}
The last equation simplifies eqs.\eqref{EV1} and \eqref{EV2}:
\begin{align}
&\rho_G(\theta ) =\frac{R}{4\kappa } \left[ 1+\frac{1-\alpha }{\sinh^2\theta}\right] - \rho_M (\theta_0)\frac{\sinh^{3/2}\theta_0}{\sinh^{3/2}\theta} \label{rhoG}\\
&  p_G (\theta) = -\frac{R}{4\kappa}\left[ 1-\frac{1-\alpha}{3\sinh^2\theta}\right] 
\end{align}
The second equation is appealing, as it says that the gravitational `dark energy' definitely exerts a constant negative pressure.
Eq.\eqref{rhoG} can be rewritten as:
\begin{align*}
\rho_G(\theta ) 
=\frac{R}{4\kappa } \left[ 1+\frac{1-\alpha}{\sinh^2\theta}-  \frac{ \alpha \Omega_M }{\Omega_R^{3/4}(\alpha-\Omega_R)^{1/4} }\frac{1}{ (\sinh\theta)^{3/2}} \right]
\end{align*}
The minimum is attained at 
\begin{gather*}
\sinh\theta_m = \frac{16}{9} \frac{\Omega_R^{3/2} (1-\alpha)^2\sqrt{\alpha-\Omega_R} }{\Omega^2_M \alpha^2}\\
\rho_G (\theta_m) = \frac{R}{4\kappa} \left[ 1-\frac{27}{256}\frac{\Omega^4_M \alpha^4}{\Omega_R^3(1-\alpha)^3 (\alpha -\Omega_R) }\right ] 
\end{gather*}
Positivity of $\rho_G(\theta_m)$ requires $\alpha =\beta\Omega_R$ with $1.0010<\beta< 9.4736$.
The maximum of $\rho_G(\theta_m)$ occurs for $\beta=4/3$.
Eq.\eqref{CON2} gives  $\cosh\theta_0 =\sqrt \beta$, then the range for $\beta $ translates into 
\begin{align}
0.0324<\theta_0 < 1.7899 \label{rangetheta}
\end{align}
Multiplication of \eqref{Hubble} by $t_0$ gives $H_0 t_0 = \frac{1}{2}\theta_0 \coth \theta_0$ i.e. the age
of the universe:
$$ t_0 = \theta_0\coth\theta_0 \times  \frac{3.086 \times 10^{19}}{2\times 67.32}  \, {\rm s}  $$ 
The range for $\theta_0$ gives a range $7.27<t_0< 13.75 $ (Gys).\\
In terms of the parameter $\beta $ or equivalenty $\theta_0$ in the range \eqref{rangetheta}, 
the fractions of the energy densities with the total energy density have evolution 
\begin{align}
&\frac{\rho_R(\theta)}{\rho(\theta)} = \Omega_R \frac{\cosh^2\theta_0}{\cosh^2\theta}\\
&\frac{\rho_M(\theta)}{\rho(\theta)} = \Omega_M \frac{\cosh^2\theta_0}{\cosh^2\theta}\frac{\sqrt{\sinh \theta}}{\sqrt{\sinh\theta_0}}\\
&\frac{\rho_G(\theta)}{\rho (\theta)} = 1-\frac{\rho_R(\theta)}{\rho(\theta)} - \frac{\rho_M(\theta)}{\rho(\theta)} 
\end{align}
\begin{figure}
\begin{center}
\includegraphics[width=7cm,clip=]{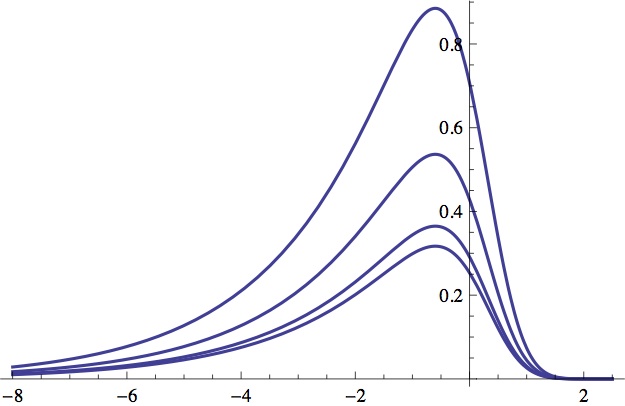}
\caption{The ratio $\rho_M/\rho$ as a function of  $\ln \theta$. From top: $\theta_0 =  1.7$, $1.3$, $0.9$ and $0.5$ (Mathematica).}
\end{center}
\end{figure}
%


\section{Conclusion}
We evaluated the Ricci tensor and the scale function of a RW space-time with constant scalar curvature and flat space.
Then, we investigated the ensuing properties of the cosmological fluid. At early and late times it features the interesting
limits $w=1/3$ and $w=-1$. To ensure positivity of all forms of energy, at small times the gravitational term $\rho_G(t)$ scales as radiation
as $t^{-2}$ and is a factor $\approx 10^4$ larger than $\rho_R(t)$. \\
After these findings, we read the scale function \eqref{at} in the book by Mukhanov \cite{Mukhanov} as characteristic 
of a RW space-time with flat space filled with radiation and cosmological constant (no matter). Here the same scale-function results 
from requiring flatness and constancy of the curvature scalar. The properties of the cosmological fluid are the outcome.

%

%
\vfill
\end{document}